# Distanced LSTM: Time-Distanced Gates in Long Short-Term Memory Models for Lung Cancer Detection

Riqiang Gao[1], Yuankai Huo[1], Shunxing Bao[1], Yucheng Tang[1], Sanja L. Antic[2], Emily S. Epstein[2], Aneri B. Balar[2], Steve Deppen[2], Alexis B. Paulson[2], Kim L. Sandler[2], Pierre P. Massion[2], Bennett A. Landman[1]

[1] Vanderbilt University, Nashville TN 37235, USA
[2] Vanderbilt University Medical Center, Nashville TN 37235, USA

**Abstract.** The field of lung nodule detection and cancer prediction has been rapidly developing with the support of large public data archives. Previous studies have largely focused cross-sectional (single) CT data. Herein, we consider longitudinal data. The Long Short-Term Memory (LSTM) model addresses learning with regularly spaced time points (i.e., equal temporal intervals). However, clinical imaging follows patient needs with often heterogeneous, irregular acquisitions. To model both regular and irregular longitudinal samples, we generalize the LSTM model with the Distanced LSTM (DLSTM) for temporally varied acquisitions. The DLSTM includes a Temporal Emphasis Model (TEM) that enables learning across regularly and irregularly sampled intervals. Briefly, (1) the temporal intervals between longitudinal scans are modeled explicitly, (2) temporally adjustable forget and input gates are introduced for irregular temporal sampling; and (3) the latest longitudinal scan has an additional emphasis term. We evaluate the DLSTM framework in three datasets including simulated data, 1794 National Lung Screening Trial (NLST) scans, and 1420 clinically acquired data with heterogeneous and irregular temporal accession. The experiments on the first two datasets demonstrate that our method achieves competitive performance on both simulated and regularly sampled datasets (e.g. improve LSTM from 0.6785 to 0.7085 on F1 score in NLST). In external validation of clinically and irregularly acquired data, the benchmarks achieved 0.8350 (CNN feature) and 0.8380 (LSTM) on area under the ROC curve (AUC) score, while the proposed DLSTM achieves 0.8905.

## 1 Introduction

Early detection of lung cancer from clinically acquired computed tomography (CT) scans are essential for lung cancer diagnosis [1]. Lung cancer detection is a binary classification (cancer or non-cancer) task from the machine learning perspective. Convolutional neural network (CNN) methods have been widely used in lung cancer detection, which typically consist of two steps: nodule detection and classification. Nodule detection detects the pulmonary nodules from a CT scan with coordinates and region of interest (e.g., [2]), while the classification assigns the nodules to be either benign or malignant categories [3], and the whole CT scan is classified as cancer when containing at least one malignant nodule. One prevalent method was proposed by Liao et al. [3], which won the Kaggle DSB2017 challenge. In this method, the pipeline was deployed on detecting top five confidence nodule regions to classify whole CT scan. The Liao et al. network focuses on a single CT scan, rather than multiple longitudinal scans.

In clinical practice, longitudinal CT scans may contain temporal relevant diagnostic information. To learn from the longitudinal scans, recurrent neural networks (RNN) have been introduced to medical image analysis when longitudinal (sequential) imaging data are available (e.g., [4]). Long Short-Term Memory (LSTM) [5] is one of the most prevalent variants of RNN, which is capable of learning both long-term and short-term dependencies between features using three gates (i.e., forget, input, and output gates). Many variants of LSTM have been proposed [6-8]. For instance, convolutional LSTM [6] is designed to deal with spatial temporal variations in images [9, 10].

In canonical LSTM, the temporal intervals between consecutive scans are equal. However, this rarely occurs in clinical practice. Temporal intervals have been modeled in LSTM for recommendation system in finance [8] and abnormality detection on 2D chest X-ray [11]. However, no previous studies have been conducted to model global temporal variations. The previous methods [8, 11] modeled the relative local time intervals between consecutive scans. However, for lung cancer detection, the last scan is typically the most informative. Therefore, we propose a new Temporal Emphasis Model (TEM) to model the global time interval between previous time points to the last scan as a global multiplicative function to input gate and forget gate, rather than a new gate as [8] or an additive term as [11].

Our contributions are: (1) this is the first study that models the time distance from last point for LSTM in lung cancer detection; (2) the novel DLSTM framework is proposed to model the temporal distance with adaptive forget gate and input gate; (3)

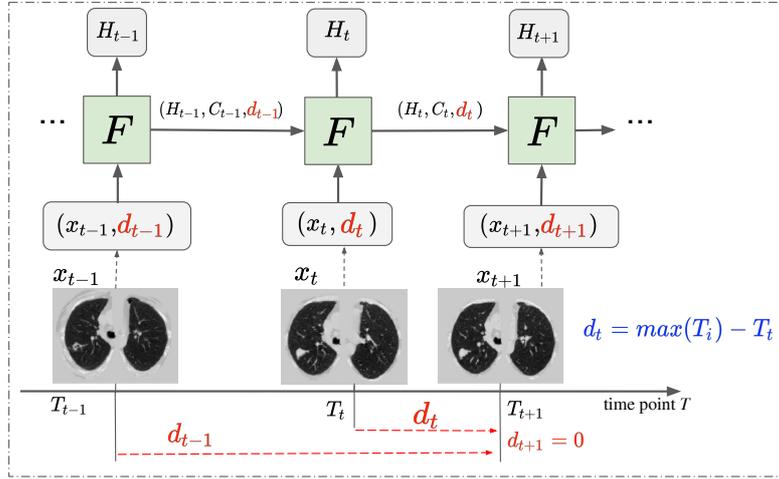

**Fig. 1.** The framework of DLSTM (three "steps" in the example). $x_t$ is the input data at time point $t$, and $d_t$ is the time distance from the time point $t$ to the latest time point. "$F$" represents the learnable DLSTM component (convolutional version in this paper). $H_t$ and $C_t$ are the hidden state and cell state, respectively. The input data, $x_t$, could be 1D, 2D, or 3D.

a toy dataset called "Tumor-CIFAR" is released to simulate dummy benign and malignant cancer on natural images. 1794 subjects from the widely used National Lung Screening Trial (NLST) [12] and 1420 subjects from two institutional cohorts are used to evaluate the methods.

## 2   Theory and Method

**Distanced LSTM -** LSTM is the most widely used RNN networks in classification or prediction upon sequential data. Standard LSTM employ three gates (i.e., forget gate $f_t$, input gate $i_t$, and output gate $o_t$) to maintain internal states (i.e., hidden state $H_t$ and cell state $C_t$). The forget gate controls the amount of information used for the current state from the previous time steps. To incorporate the "distance attribute" to LSTM, we multiply a Temporal Emphasis Model (TEM) $D(d_t, a, c)$ as a multiplicative function to the forget gate and the input gate with learnable parameters (Figure 1).

Briefly, our DLSTM is defined by following the terms and variables in [6]:

$$\begin{aligned} i_t &= D(d_t, a, c) \cdot \sigma(W_{xi} * X_t + W_{hi} * H_{t-1} + W_{ci} \circ C_{t-1} + b_i) \\ f_t &= D(d_{t-1}, a, c) \cdot \sigma(W_{xf} * X_t + W_{hf} * H_{t-1} + W_{cf} \circ C_{t-1} + b_f) \\ C_t &= f_t \circ C_{t-1} + i_t \circ tanh(W_{xc} * X_t + W_{hc} * H_{t-1} + b_i) \\ o_t &= \sigma(W_{xo} * X_t + W_{ho} * H_{t-1} + W_{co} \circ C_t + b_o) \\ H_t &= o_t \circ tanh(C_t) \end{aligned} \quad (1)$$

where $x_t$ is the input data of time point $t$, $d_t$ is the global time distance from any $x_t$ to the latest scan, $W$ and $b$ are the learnable parameters, and "*" and "∘" denotes the convolution operator and Hadamard product respectively. Different from canonical LSTM, the TEM function is introduced in the proposed DLSTM as

$$D(d_t, a, c) = a \cdot e^{-c \cdot d_t} \quad (2)$$

where $a$ and $c$ are positive learnable parameters. Different from tLSTM [11], which introduced an additive term to model local relative time interval between scans, the proposed DLSTM introduces the TEM function as a global multiplicative function to model the time interval (distance) from each scan to the last scan. Using TEM in Eq. (1), both the forget gate $f_t$ and input gate $i_t$ are weakened if the input scan is far from the last scan. Note the "LSTM" represents the convolutional version in this paper.

## 3   Experiment Design and Results

We include both simulation (Tumor-CIFAR) and empirical validations (NLST and clinical data from two in-house projects, see Table 1) to validate the baseline methods and the proposed method. Firstly, to test if our algorithm can handle the time-interval distances effectively, we introduce the synthetic dataset: Tumor-CIFAR.

In Tumor-CIFAR, we show the test results with a training/validation/test split (Figure 2). We perform three different validations on lung datasets: (1) cross-validation on NLST with longitudinal data (Table 2); (2) cross-validation on clinical data with both cross-sectional and longitudinal scans (Table 3); and (3) external-validation on longitudinal scans (train and validation on NLST and test result on clinical data, Table 4).

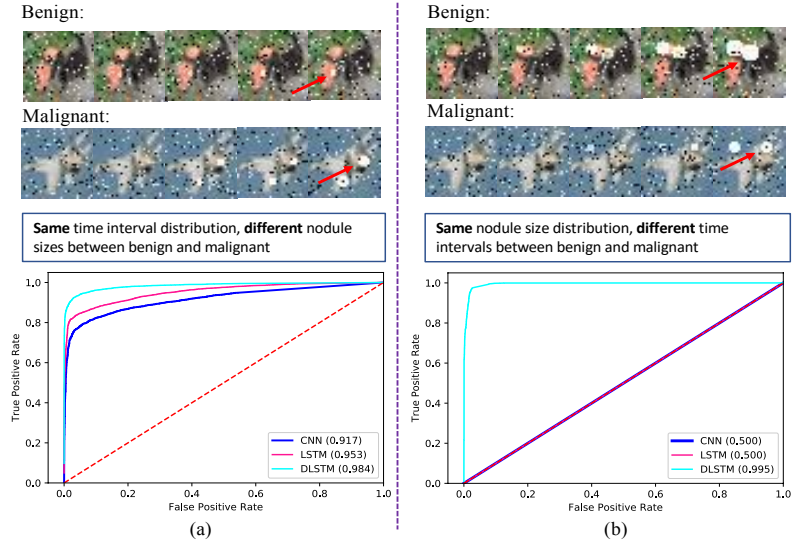

**Fig. 2.** The receiver operating characteristic (ROC) curves of Tumor-CIFAR. The left panels simulate the situation that the images are sampled with the same interval distribution, while the right panels are sampled with the same size distribution. The upper panels show the examples of images in Tumor-CIFAR. The noise (white and black dots) are added, while the dummy nodules are shown as white blobs (some are indicated by red arrows). The lower panels show the Area Under the Curve of ROC (AUC) values of different methods.

### 3.1 Simulation: Tumor-CIFAR

**Data.** Based on [13], the growth speed of malignant nodules is approximate three times faster compared with benign ones. To incorporate temporal variations in the simulation, we add dummy nodules on CIFAR10 [14] with different growth rate for benign and malignant nodules (malignant nodules grow three times faster than benign ones). Two cases are simulated: image samples with the same "interval distribution" (Figure 2a), or with the same nodule "size distribution" (Figure 2b). The same interval distribution indicates intervals follow the same Gaussian distribution. The same nodule size distribution represents the growth rate of nodules follow the same Gaussian distribution (simulation code, detail descriptions and more image examples are publicly available at https://github.com/MASILab/tumor-cifar).

There are 5,000 samples in the training set and 1,000 samples in the testing set. Cancer prevalence was 50% in each dataset. Each sample is simulated with five different time points. The training/validation/test split is 40k/10k/10k.

**Experimental Design.** The base network structure (CNN in Figure 1) is employed from the official PyTorch 0.41 [15] example for MNIST (we call it "ToyNet"). The ToyNet is composed of two convolutional layers (the second with a 2D dropout) and followed by two fully connected layers along with a 1D dropout in the middle. "LSTM" and "DLSTM" in Figure 2 represents a 2D convolutional LSTM component and 2D convolutional of our proposed DLSTM component is stacked in the beginning of the "ToyNet", respectively. The maximum training epoch number is 100. The initial learning rate set to 0.01 and is multiplied by 0.4 at $50^{th}$, $70^{th}$ and $80^{th}$ epoch.

**Table 1.** Demographic distribution in our experiments

| Lung Data Source | NLST | MCL | VLSP |
|---|---|---|---|
| Total Subject | 1794 | 567 | 853 |
| Longitudinal Subject | 1794 | 105 | 370 |
| Cancer Frequency (%) | 40.35 | 68.57 | 2.00 |
| Gender (male, %) | 59.59 | 58.92 | 54.87 |

**Results.** For the same time interval distribution (Figure 2a), the LSTM achieves higher performance compared with baseline CNN method, while the DLSTM works even better. This task is relatively easy since the malignant nodules clearly grow faster compared benign nodules. However, if we control the sampling strategy to guarantee the same nodule size for corresponding samples (Figure 2b), the task becomes challenging if the time intervals are not modeled in the network design since the nodules are now having the same size. In this case, the CNN and LSTM only achieve 0.5 AUC values, while our DLSTM is able to almost perfectly capture the temporal variations with an AUC value of 0.995.

### 3.2 Empirical Validation on CT

**Data.** The National Lung Screening Trial (NLST) [12] is a large-scale randomized controlled trial for early diagnosis of lung cancer study with low-dose CT screening exams publicly available. We obtain a subset (1794 subjects) from NLST, which contains all longitudinal scans with "follow-up confirmed lung cancer", as well as a random subset of all "follow-up confirmed not lung cancer" scans (Table 1). One in-house dataset combines two clinical lung sets Molecular Characterization Laboratories (MCL, https://mcl.nci.nih.gov) and Vanderbilt Lung Screening Program (VLSP, https://www.vumc.org/radiology/lung) which is also evaluated by our algorithm. These data are used in de-identified form under internal review board supervision.

**Experimental Design.** The DLSTM can be trained in an end-to-end network (simulation experiments in Section 3.1) or as lightweight post-processing manner. In this section, we evaluate the proposed DLSTM as a post processing network for the imaging features extracted from Liao et al. [3]. We compare the DLSTM with a recently proposed benchmark tLSTM [11], which models the relative time interval as an additive term. Five highest risk regions (possible nodules) for each scan are detected by [3], and the feature dimension for each region is 64, then the scan-level feature is achieved by concatenating region features as 5×64 inputs. For a fair comparison, the same features are provided to the networks Multi-channel CNN (MC-CNN), LSTM, tLSTM, and DLSTM, with 1D convolutional layer of 5 kernel size. MC-CNN concatenates multi-scan features in the "channel" dimension. The maximum training epoch number is 100, the initial learning rate is set to 0.01 and multiplied by 0.4 at the $50^{th}$, $70^{th}$, and $80^{th}$ epoch. Since most of the longitudinal lung CT scans contain two time points, we

6evaluate the MC-CNN, LSTM, tLSTM and DLSTM with two time points ("2 steps") in this study (the last two points are picked if the patient with more than two scans).

The "Ori CNN" in Tables 2-4 represents the results obtained by open source code and trained model of [3]. If there is on special explanation, our results are reported at subject-level rather than scan-level, and the "Ori CNN" reports the performance of the latest scan of patients.

**Preprocessing**. Our preprocessing follows Liao et al. [3]. We resample the 3D volume to $1 \times 1 \times 1$ mm isotropic resolution. The lung CT scan is segmented using (https://github.com/lfz/DSB2017) from the original CT volume and the non-lung regions are zero-padded to Hounsfield unit score of 170. Then, the 3D volumes are resized to 128×128×128 to use pre-trained model for extracting image features.

**Results: Cross-validation on longitudinal scans.** Table 2 shows the five-fold cross-validation results on 1794 longitudinal subjects from the NLST dataset. All the training and validation data are longitudinal (with "2 steps").

**Results: Cross-validation on combining cross-sectional and longitudinal scans.** More than half of the patients only have cross-sectional CT (single time point) scans from clinical projects (see Table 1). Therefore, we evaluate the proposed method as well as the baseline methods on the entire clinical cohorts with both longitudinal and cross-sectional testing with cross-validation on all 1420 subjects by duplicating scans for subjects with only one scan to 2 steps. Table 3 indicates the five-fold cross-validation results on the clinical data. As for tLSTM [11] and the proposed DLSTM, we set the time interval and time distance to be zero for cross-sectional scans, respectively.

Table 2. Experimental results on NLST dataset (%, average (std) of cross-validation)

| Method | Accuracy | AUC | F1 | Recall | Precision |
|---|---|---|---|---|---|
| Ori CNN [3] | 71.94(2.07) | 74.18(2.11) | 52.18(2.83) | 38.07(2.63) | 83.24(4.24) |
| MC-CNN | 73.26(3.10) | 77.96(0.98) | 59.39(3.70) | 47.91(4.87) | 78.62(3.09) |
| LSTM [5, 6] | 77.05(1.46) | 80.84(1.20) | 67.85(2.41) | 59.92(4.43) | 78.68(3.32) |
| tLSTM [11] | 77.37(2.97) | 80.80(1.45) | 67.47(3.58) | 58.65(5.12) | 79.81(3.34) |
| DLSTM(ours) | **78.96(1.57)** | **82.55(1.31)** | **70.85(1.82)** | **61.61(2.01)** | **83.38(4.34)** |

\* The AUC represents Area Under the Curve of receiver operating characteristic, and best result is shown in **bold** (also used in the following).

Table 3. Experimental results on clinical datasets (%, average (std) of cross-validation)

| Method | Accuracy | AUC | F1 | Recall | Precision |
|---|---|---|---|---|---|
| Ori CNN [3] | 84.80(2.43) | 89.00(1.65) | 70.29(4.26) | 63.46(3.51) | 78.83(5.70) |
| MC-CNN | 84.51(1.29) | 90.85(1.13) | 70.55(1.29) | 62.85(1.53) | 80.84(4.42) |
| LSTM [5, 6] | 86.27(1.29) | 90.27(1.15) | 74.17(2.47) | 69.73(2.62) | 79.56(5.69) |
| tLSTM [11] | 86.42(1.48) | 91.06(1.48) | 74.36(1.99) | 68.55(1.55) | **81.49(5.28)** |
| DLSTM(ours) | **86.97(1.45)** | **91.17(1.53)** | **76.11(2.68)** | **72.71(2.38)** | 80.04(5.18) |

Table 4. Experimental results on cross-dataset test (external-validation)

| Method | Accuracy | AUC | F1 | Recall | Precision |
|---|---|---|---|---|---|
| Train and Test both on longitudinal subjects | | | | | |
| Ori CNN (all scans) | 0.8342 | 0.8350 | 0.5253 | 0.4577 | 0.6266 |
| Ori CNN [3] | 0.8758 | 0.8510 | 0.5931 | 0.5513 | **0.6418** |
| MC-CNN | 0.8589 | 0.7654 | 0.5621 | 0.5513 | 0.5733 |
| LSTM [5, 6] | 0.8589 | 0.8380 | 0.5732 | 0.5769 | 0.5692 |
| tLSTM [11] | 0.8673 | 0.8869 | 0.6631 | **0.7949** | 0.5688 |
| DLSTM(ours) | **0.8863** | **0.8905** | **0.6824** | 0.7436 | 0.6304 |

**Results: External-validation on longitudinal scans.** We directly apply the trained models from NLST to the in-house subjects as external validation, without any further parameter tuning (Table 4). Note that the longitudinal data are regularly sampled in NLST while the clinical datasets are irregularly acquired. The final predicted cancer probability is the average of five models trained on five-folds of NLST. The "Ori CNN (all scans)" in Table 4 represents the scan-level results of all scans from longitudinal subjects.

**Analyses:** In both public dataset NLST and our private datasets, the proposed DLSTM achieves competitive results in accuracy, AUC, F1, recall and precision. For example, the proposed DLSTM improves the conventional LSTM on F1 score from 0.6785 to 0.7085 (Table 2, NLST dataset), and from 0.7417 to 0.7611 (Table 3, clinical datasets). External validation experiments indicate the generalization ability of the proposed method.

In the external validation, (1) the latest scans achieve higher performance compare with longitudinal scans, which indicates that emphasis on latest longitudinal scan in our DLSTM is meaningful. (2) the algorithms with time information (tLSTM and the proposed DLSTM) outperform those methods without temporal emphasis when the test dataset is irregularly sampled.

## 4    Conclusion and Discussion

In this paper, we propose a novel DLSTM method to model the global temporal intervals between longitudinal CT scans for lung cancer detection. Our method has been validated using both simulations on Tumor-CIFAR, empirical validations on 1794 NLST and 1420 clinically subjects. From cross-validation and external-validation, the proposed DLSTM method achieves generally superior performance compared with baseline methods. Meanwhile, the Tumor-CIFAR dataset is publicly available.

**Acknowledgments:** This research was supported by NSF CAREER 1452485, 5R21 EY024036, R01 EB017230. This study was supported in part by a UO1 CA196405 to Massion. This study was in part using the resources of the Advanced Computing Center for Research and Education (ACCRE) at Vanderbilt University, Nashville, TN. This project was supported in part by the National Center for Research Resources, Grant UL1 RR024975-01, and is now at the National Center for Advancing Translational Sciences, Grant 2 UL1 TR000445-06. We gratefully acknowledge the support of NVIDIA Corporation with the donation of the Titan X Pascal GPU used for this research. The de-identified imaging dataset(s) used for the analysis described were obtained from ImageVU, a research resource supported by the VICTR CTSA award (ULTR000445 from NCATS/NIH), Vanderbilt University Medical Center institutional funding and Patient-Centered Outcomes Research Institute (PCORI; contract CDRN-1306-04869). This research was also supported by SPORE in Lung grant (P50 CA058187), University of Colorado SPORE program, and the Vanderbilt-Ingram Cancer Center.